\documentclass[12pt]{article}

\usepackage{siunitx}
\usepackage{times}
\usepackage[version=4]{mhchem}
\usepackage{graphicx}
\usepackage{lipsum}
\usepackage{bm}
\usepackage{color}
\usepackage[pdftex]{hyperref}
\hypersetup{
	colorlinks=true,       % false: boxed links; true: colored links
    linkcolor=blue,          % color of internal links (change box color with linkbordercolor)
    citecolor=black,        % color of links to bibliography
    filecolor=yellow,      % color of file links
    urlcolor=blue           % color of external links
    }
\title{Atomic reconstruction in twisted bilayers of transition metal dichalcogenides} 

\author
{Astrid Weston,$^{1,2}$ Yichao Zou,$^{2,3}$ Vladimir Enaldiev,$^{1,2,4}$ \\
Alex Summerfield,$^{1,2}$ Nicholas Clark,$^{2,3}$ Viktor Z\'olyomi,$^{1,2}$ \\
Abigail Graham,$^{5}$ Celal Yelgel,$^{1,2}$ Samuel Magorrian,$^{1,2}$ \\
Mingwei Zhou,$^{1,2}$  Johanna Zultak,$^{1,2}$ David Hopkinson,$^{2,3}$ \\
Alexei Barinov,$^{6}$ Thomas Bointon,$^{2}$ Andrey Kretinin,$^{2,3}$ \\ 
Neil R. Wilson,$^{5}$ Peter H. Beton,$^{7}$  Vladimir I. Fal'ko,$^{1,2,8*}$ \\
Sarah J. Haigh,$^{2,3*}$ Roman Gorbachev$^{1,2,8*}$ 
\\
\\
\normalsize{$^{1}$Department of Physics and Astronomy, University of Manchester, Oxford Road,}\\
\normalsize{ Manchester, M13 9PL, UK}\\
\normalsize{$^{2}$National Graphene Institute, University of Manchester, Oxford Road,}\\
\normalsize{ Manchester, M13 9PL, UK}\\
\normalsize{$^{3}$Department of Materials, University of Manchester, Oxford Road,}\\
\normalsize{ Manchester, M13 9PL, UK}\\
\normalsize{$^{4}$Kotel'nikov Institute of Radio-engineering and Electronics,} \\ 
\normalsize{Russian Academy of Sciences, 11-7 Mokhovaya St, Moscow, 125009 Russia}\\
\normalsize{$^{5}$Department of Physics, University of Warwick, Coventry, CV4 7AL, UK}\\
\normalsize{$^{6}$Elettra - Sincrotrone Trieste, S.C.p.A., Basovizza (TS), 34149, Italy}\\
\normalsize{$^{7}$School of Physics and Astronomy, University of Nottingham, Nottingham, NG7 2RD, UK}\\
\normalsize{$^{8}$Henry Royce Institute for Advanced Materials, University of Manchester,}\\
\normalsize{Oxford Road, Manchester, M13 9PL, UK}\\
\\
\normalsize{ E-mail: vladimir.falko@manchester.ac.uk, sarah.haigh@manchester.ac.uk,}
\\
\normalsize{roman@manchester.ac.uk.}
}

\date{}

\begin{document} 

\nopagebreak
\maketitle 

\newpage
\begin{small} \vspace*{-62pt}
\noindent This manuscript has been substantially updated and includes additional experimental results on piezoelectric textures created by the lattice reconstruction. 
Latest version is available at:  \url{https://www.nature.com/articles/s41565-020-0682-9} 
\end{small}
%\newpage 

\begin{abstract}
Van der Waals heterostructures form a massive interdisciplinary research field, fueled by the rich material science opportunities presented by layer assembly of artificial solids with controlled composition, order and relative rotation of adjacent atomic planes. Here we use atomic resolution transmission electron microscopy and multiscale modeling to show that the lattice of MoS$_2$ and WS$_2$ bilayers twisted to a small angle, $\theta<3^{\circ}$, reconstructs into energetically favorable stacking domains separated by a network of stacking faults. For crystal alignments close to 3R stacking, a tessellated pattern of mirror reflected triangular 3R domains emerges, separated by a network of partial dislocations which persist to the smallest twist angles. Scanning tunneling measurements show that the electronic properties of those 3R domains appear qualitatively different from 2H TMDs, featuring layer-polarized conduction band states caused by lack of both inversion and mirror symmetry. In contrast, for alignments close to 2H stacking, stable 2H domains dominate, with nuclei of an earlier unnoticed metastable phase limited to $\sim$ \SI{5}{\nm} in size. This appears as a kagome-like pattern at $\theta\sim 1^{\circ}$, transitioning at $\theta\rightarrow 0$ to a hexagonal array of screw dislocations separating large-area 2H domains.
\end{abstract}

Moir\'e superlattices generated at the interfaces of two-dimensional crystals with a small relative twist or lattice mismatch have proved to be a powerful tool for controlling the electronic and optical properties of van der Waals heterostructures. In crystallographically aligned graphene/boron nitride (hBN) heterostructures, moir\'e superlattices lead to the formation of mini-bands for electrons and the Hofstadter butterfly effect \cite{Ponomarenko2013,Dean2013}.  Moir\'e superlattices in twisted bilayer graphene have already produced superconductivity \cite{Cao2018}, a Mott insulator-like state \cite{Cao22018}, and helical networks of topological edge states \cite{Yin2016,Huang2018}. Significant breakthroughs have also been achieved in understanding the potential of moir\'e superlattices in transition metal dichalcogenide (TMD) twisted bilayers. Localisation of excitons by stacking-dependent modulation of the bandgap has been observed in WSe$_2$/MoSe$_2$ \cite{Seyler2019,Tran2019} and WSe$_2$/WS$_2$ \cite{Jin2019} hetero-bilayer structures, while MoSe$_2$/WS$_2$ heterostructures have demonstrated the formation of minibands of resonantly hybridized excitons, promoted by the close proximity of the conduction band edges in the two compounds \cite{Alexeev2019}.

The electronic structure of bilayers is known to depend critically on the local atomic stacking configuration. Despite its relative weakness, the coupling between neighboring van der Waals layers induces an atomic lattice reconstruction in the constitutent crystals, already observed in both graphene/hBN and twisted graphene bilayers \cite{Woods2014,Yoo2019}. For graphene bilayers with a small misalignment angle, $\theta \le 1^{\circ}$, a monotonic variation of the interlayer lattice registry transforms into a pattern of triangular Bernal-stacked domains separated by a network of stacking faults observed from the diffraction contrast present in low magnification transmission electron microscopy (TEM) images \cite{Alden2013, Zhang2018, Butz2014}. The presence of lattice reconstruction has also been suggested \cite{Naik2019, Naik2018, Carr2018} in TMD bilayers, but, despite enormous scientific interest in this system, an experimental study of lattice reconstruction in twisted TMDs is still missing. Here we use atomic resolution scanning transmission electron microscopy (STEM), accompanied by first-principle and multiscale modeling, to study how a network of domains with strong local commensuration develops at small twist angles in suspended WS$_2$ and MoS$_2$ homo- and hetero-bilayers. We find that, while lattice reconstruction in twisted bilayers close to 3R-type (parallel alignment) is morphologically similar to graphene, twisted homo- and hetero-bilayers of TMDs offer a broader diversity of physical properties prescribed by the lack of inversion symmetry in the constituent layers. This makes the resulting structure qualitatively different in twisted TMD bilayers not just morphologically, but also electronically. In particular, lattice reconstruction for P-oriented bilayers creates conditions for the formation of sizable domains of 3R stacking (which is rarely found in bulk TMD crystals \cite{Suzuki2014, Ubrig2017}) and features intrinsic asymmetry of electronic wavefunctions, which has not been observed before in 2D materials. The latter observation was made using high-resolution conductive AFM tunneling experiments in agreement with Density Functional Theory (DFT) modeling. Moreover, strikingly different patterns emerge in 2H-type (anti-parallel) bilayers, where we find kagome-like dislocation network.

\begin{figure}
\centering
\includegraphics[width=\textwidth/3]{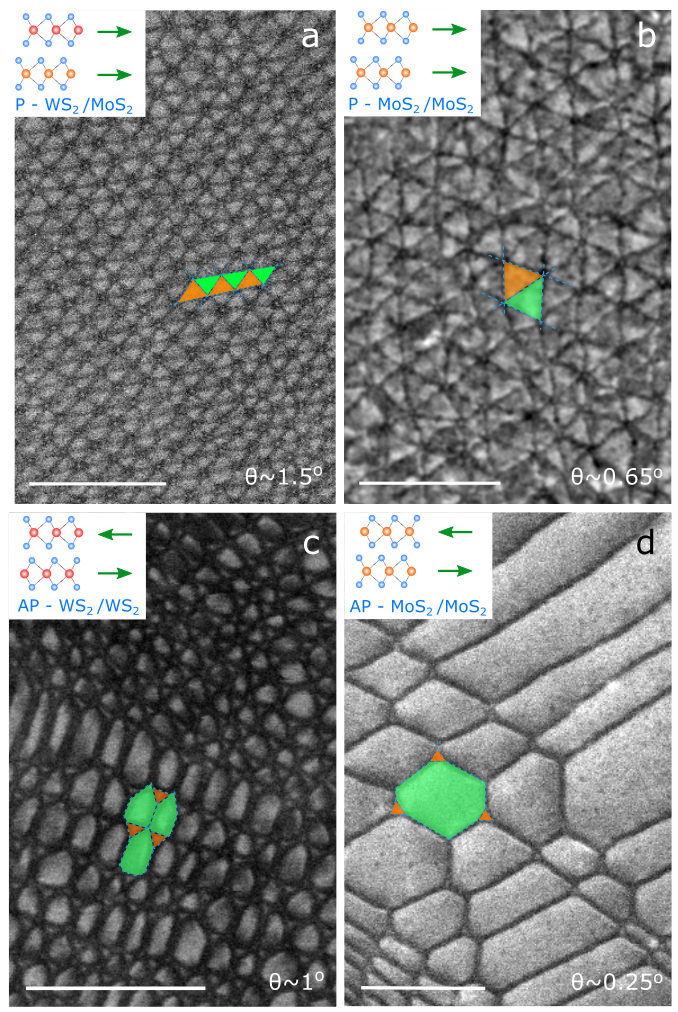}
\caption{\label{fig:I}\textbf{Lattice domains in homo- and hetero-bilayers of MoS$_2$ and WS$_2$.} 
Low magnification ADF STEM image of periodic domain array for various twist angles, scale bar \SI{100}{nm}. Atomic schematics insets illustrate parallel and antiparallel orientation of the lattices disregarding the local stacking configuration.}
\end{figure}

Bulk TMD crystals are composed of regularly stacked monolayers, each consisting of a triangular sublayer of metal atoms at the mirror-symmetry plane between two identical triangular chalcogen sublayers. In the 2H TMD polytype (space group P63/mmc), all metal (M) atoms are vertically aligned with the chalcogen (X) atoms in the nearest neighbor layers. In contrast, in the 3R polytype (space group R3m), chalcogens are aligned with the empty centers of the hexagons in the layers above and below. To recreate these polytypes in twisted homo-bilayers, we employ the tear-and-stamp transfer technique \cite{Kim2016}, where one half of an exfoliated monolayer flake is picked up and deposited onto the other half. In this case, perfect parallel (P) alignment of the crystalline lattices would result in 3R stacking, while an exact $180^{\circ}$ rotation (anti-parallel, AP) would produce 2H stacking order. A small deviation from a perfect P or AP alignment (to an angle $\theta \sim 0.5^{\circ} - 3^{\circ}$) gives rise to a moir\'e pattern with a period, $\ell=a/2\sin\frac{\theta}{2}$, much larger than the in-plane lattice constant, $a$, of the TMD. To fabricate hetero-bilayers with P or AP alignment we use a similar approach, where the parallel edges of the crystals are aligned during the transfer process \cite{Ponomarenko2013}. The assembled stacks are then transferred onto custom-made support grids (see SI S1) to allow high-resolution STEM imaging. For scanning tunneling measurements, the assembled bilayers are transferred onto multilayer graphite pre-exfoliated onto an oxidized silicon substrate. 

The presence of domains in the resulting moir\'e patterns is clearly seen from the diffraction contrast in low-magnification STEM images, Fig.1, for both homo- and hetero-bilayers of MoS$_2$ and WS$_2$. For P layer alignment, we observe triangular domains with boundaries visible as thin lines of darker contrast. These domains retain periodicity over hundreds of micrometers, distorted by global strain unavoidable in suspended 2D crystals (see SI S2). For AP orientations (close to 2H), we consistently observe a kagome-like pattern where the domain structure is dominated by hexagonally shaped regions (examples marked green in Fig.1c,d) with small ($\sim$5\,nm) triangular seeds of an alternative stacking sequence (marked orange) occupying every other vertex. Similar domain patterns are visible in AFM topography and friction scanning measurements obtained on twisted TMD bilayers deposited on a thick graphite crystal pre-exfoliated onto an oxidized silicon wafer (see SI S6).

\begin{figure}
\centering
\includegraphics[width=\textwidth]{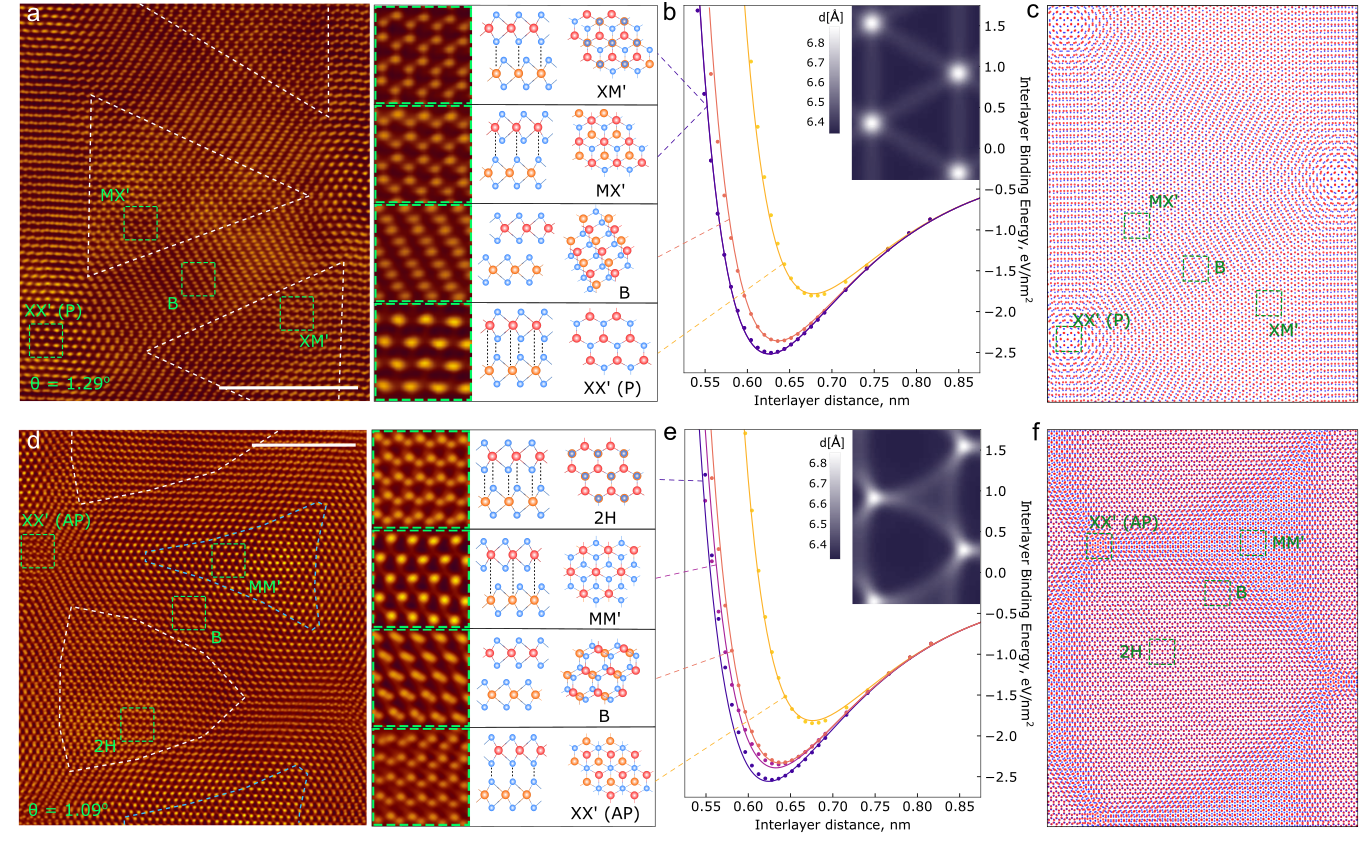}
\caption{\label{fig:II}\textbf{Lattice reconstruction in twisted WS$_2$ homo-bilayers.} Filtered annular dark field (ADF) STEM image of suspended twisted WS$_2$ bilayer for parallel (\textbf{a}) and antiparallel (\textbf{d}) alignment with commensurate domains highlighted by dashed lines. Selected areas are shown magnified on the left with corresponding atomic schematics, the scale bars are 5\,nm. Interlayer binding energy density computed using vdW-DFT for the indicated stacking sequence as a function of interlayer distance (centre to centre of each layer) for P (\textbf{b}) and AP (\textbf{e}) orientations with a fit to Eq.(1) in solid lines with $C_4=0.137976$ eV$\cdot$nm$^2$, $C_8=0.159961$ eV$\cdot$nm$^6$, $C_{12}=-0.020753$ eV$\cdot$nm$^{10}$, $A_1=8.45716\times 10^7$ eV/nm$^2$, $A_2=7.0214\times 10^4$ eV/nm$^2$, $q=30.877$ nm$^{-1}$, $\varepsilon=213$ eV/nm$^4$, $d_0=0.651$ nm. Calculated lattice reconstruction for $\theta=1.29^{\circ}$ in P-WS$_2$ \textbf{(c)} and $\theta=1.09^{\circ}$ in AP-WS$_2$ \textbf{(f)} with 3R (MX$^\prime$, XM$^\prime$), MM$^\prime$ and 2H regions marked in the appropriate positions. Boundary regions labeled ``B'' in a, d, c, f have continually varying incommensurate stacking. Corresponding interlayer distance maps are shown in the insets of (b) and (e).}
\end{figure}

In Fig.2a,d we show atomic resolution STEM images of WS$_2$ homo-bilayers with $\theta = 1.29^{\circ}$ P and $\theta = 1.09^{\circ}$ AP alignments, respectively. Both images show multiple different local stacking configurations, as exemplified in the magnified images and schematics to the right of the main image. Both P and AP orientations show extended domains of commensurate (3R or 2H) stacking occupying a significant portion of the moir\'e supercell area, achieved by a local $\pm\theta/2$-rotation of the lattices around the domain center.  In the P-bilayers, Fig.2a, the nearly triangular domains ($\sim 8$\,nm long) correspond to the commensurate MX$^\prime$ and XM$^\prime$ stacking (where the top monolayer metal atom sits directly above the bottom monolayer chalcogen and {\it vice versa}, as in the 3R polytype). Domain boundaries are $\sim$\SI{3}{\nm} wide and show a continuous change in atomic registry, with the boundaries converging at points of XX$^\prime$ stacking (where neighboring chalcogens in the two monolayers are located directly above each other). In contrast, for AP orientation bilayers the 2H-stacking domains dominate over a metastable commensurate phase corresponding to MM$^{\prime}$ stacking (where the metal atoms of the top and bottom monolayers are located directly above each other). The two phases are separated by domain boundaries (width ~ $\sim$ \SI{3.5}{\nm}) converging at XX$^{\prime}$ points surrounded by a region without any discernible commensurability. 

To understand the energetics of domain formation we computed interlayer binding energy densities, $W(\bm{r}_0,d)$, for the various stacking configurations realized locally across the moir\'e supercell, encoded by a mutual shift $\bm{r}_0$ ($|\bm{r}_0|<a$) of sulfur atom positions in the two layers (Fig.2b,e) and interlayer distance $d$. The highest energy in both P and AP orientations is found for XX$^\prime$ stacking ($\bm{r}_0=0$). For P-type bilayers the most energetically favorable stacking is XM$^\prime$ or MX$^\prime$ (corresponding to $\bm{r}_0=(0,a/\sqrt{3})$ and $\bm{r}_0=(0,-a/\sqrt{3})$, respectively), which are symmetric with respect to the basal plane reflection and have the same energy. For the AP case, perfect 2H stacking has the lowest energy, with MM$^\prime$ stacking being less favourable, which is reflected in the disparity of their sizes evident in Fig.2d.

The computed values of $W(\bm{r}_0,d)$, shown in Fig.2b and 2e, are in a qualitative agreement with the earlier studies \cite{Naik2019, Naik2018} and are well-described by an interpolation formula,     
\begin{equation}\label{adh_en}
W_{AP/P}(\bm{r}_0,d) = f(d) + \sum_{n=1}^{3}\left[
A_1 e^{-qd}\cos\left(\bm{G}_n\bm{r}_0\right)+A_2 e^{-G d}%\sum_{n=1}^{3}
\sin\left(\bm{G}_n\bm{r}_0+\varphi_{AP/P}\right)\right]. 
\end{equation}
Here, $f(d)=-\sum_{i=1}^{3}C_{4i}/d^{4i}\sim -C_{4}/d^4$ (at long distances) accounts for the asymptotic behavior of the van der Waals attraction; $\bm{G}_{1,2,3}$ are the reciprocal lattice vectors in the first star of the TMD monolayer, $G=|\bm{G}_{1,2,3}|$, $\varphi_{AP}=0$ and $\varphi_P=\pi/2$, and $q$ characterizes tunneling between chalcogens. The values of the parameters in Eq. (\ref{adh_en}) fitted to the DFT-vdW data for the WS$_2$ homo-bilayer are listed in the caption of Fig.2 (for the MoS$_2$ homo-bilayer and MoS$_2$/WS$_2$ hetero-bilayer see Table S1 in SI). We note that vdW-DFT overestimates the equilibrium layer spacing as compared to the experimentally measured bulk value, but the shape of $W(\bm{r}_0,d)$ as a function of $d$ is believed to be accurate \cite{klimevs2011van}. Therefore, in the following we only analyze relative changes in the interlayer distance using a quadratic expansion for $f\approx f(d_0)+\varepsilon(d-d_0)^2$ about minimum, $d_0$, and linear expansions (in $d-d_0$) for the exponential factors. In order to test the validity of our model we have described the layer breathing mode in a 2H WS$_2$ (MoS$_2$) bilayer, and find our vdW-DFT parametrized Eq.(1) gives the phonon frequency $\omega_{LBM}\approx30$ cm$^{-1}$ ($\approx 37$\,cm$^{-1}$), in good agreement with the experimentally measured values of $\omega_{LBM}=31 - 33$ cm$^{-1}$ ($39 - 41$\,cm$^{-1}$) \cite{yang2017excitation,chen2015helicity}. 

To achieve a quantitative description of the atomic reconstruction of the observed twisted TMD bilayers as a function of twist angle, we employ multiscale modeling, similar to the approach used in \cite{Carr2018}. For two rigid monolayers, twisted to a small angle, $\theta\le 5^{\circ}$, the interlayer shift, $\bm{r}_0\approx \theta \hat{z}\times \bm{r}$, periodically repeats the same local stacking configurations in consecutive moir\'e superlattice unit cells. Lattice relaxation introduces an additional shift such that, $\bm{r}_0 \approx \theta \hat{z}\times \bm{r} + \bm{u}_{tb}$, where relative displacements are determined by the strain distribution, $u_{ij}^{t/b}(\bm{r})$, in the top (t) / bottom (b) layers, and  $\partial_ju_{i,tb}+\partial_iu_{j,tb} = 2\left(u_{ij}^{t}-u_{ij}^{b}\right)$. We also take into account out-of-plane distortions of the layers by finding the optimal interlayer distance for the various local stacking configurations using the above-described expansion around $d_0$, with a height profile illustrated in the insets in Fig.2b,e. At this point, we neglect the energy of bending deformation, estimated to be much less than other energy scales in the problem, especially for longer-period moire structures, as the energy cost of bending scales as inverse 4th order of the moir\'e pattern period. Note that AFM topography of a twisted bilayer (presented in SI S6 for a MoS$_2$ homo-bilayer) reveals height variation at the boundaries, e.g. between XM$^\prime$ and MX$^\prime$ domains, where the interlayer distance is expected to swell, according to eq.(1) and Fig.2b. Note that such adjustment of the interlayer distance to the local stacking promotes the lattice reconstruction, as compared to the predictions made in \cite{Carr2018}. 

Hence, we calculate the expected lattice reconstruction from the competition between elastic and adhesion energies,
\begin{eqnarray}\label{functional}
W_{AP/P}^{\tiny\mbox{total}}=\int d^2\bm{r}\Big\{\sum_{l={t,b}}\left[\left(\lambda/2\right)\left(u_{l,ii}\right)^{2} + \mu u_{l,ij}^{2}\right] -\varepsilon\widetilde{d}^2_{AP/P}(\bm{r},\bm{u}_{tb}) \qquad \qquad \qquad \qquad  \qquad \\
+ \sum_{n=1}^{3}\left[A_{1}e^{-qd_0}\cos\left(\bm{g}_n\bm{r} + \bm{G}_n\bm{u}_{tb}\right)+A_2e^{-Gd_0}\sin\left(\bm{g}_n\bm{r} + \bm{G}_n\bm{u}_{tb}+\varphi_{AP/P}\right)\right]\Big\},  \nonumber
\end{eqnarray}
\begin{eqnarray}
\widetilde{d}_{AP/P}(\bm{r},\bm{u}_{tb})=\frac{1}{2\varepsilon}\sum_{n=1}^{3}[qA_{1}e^{-qd_0}\cos(\bm{g}_n\bm{r} + \bm{G}_n\bm{u}_{tb})+ \qquad  \qquad \qquad \qquad \qquad \nonumber \\
GA_{2}e^{-Gd_0}\sin\left(\bm{g}_n\bm{r} + \bm{G}_n\bm{u}_{tb}+\varphi_{AP/P}\right)]. \nonumber
\end{eqnarray}
Here, $\lambda$ and $\mu$ are the first Lam\'e parameter and shear modulus of a TMD monolayer,  $\bm{g}_n=\theta \hat{z}\times \bm{G}_n $ are the supercell reciprocal vectors. To find the optimal $\bm{u}_{tb}$, we expand it in a Fourier series over $\bm{g}_n$ and self-consistently solve a system of non-linear equations for their Fourier amplitudes (see S9 SI). 

The results of such simulations for P and AP bilayers of WS$_2$ with $\theta=1.29^{\circ}/1.09^{\circ}$ are shown in Fig.2c and f for comparison with the local lattice reconstruction determined using high-resolution STEM, Fig.2a and d. The sizes and local stacking order of the computed domains show excellent agreement with the experimental data. Domain boundaries can also be identified by their larger interlayer spacing using the maps of out-of-plane displacements, inset in Fig.2b and e. In Fig.3a we present an example of a detailed comparison between the computed lattice reconstruction and STEM images of a AP aligned WS$_2$ homo-bilayer with $\theta=1.24^{\circ}$. The contrast in these STEM images is dominated by the strong electron scattering of the heavier metal atoms, so for visualisation purposes the theoretically computed reconstructed WS$_2$ bilayer lattice is represented only by the tungsten sites in both top (blue) and bottom (red) layers. When placed over a greyscale map of STEM intensity in Fig.3a, it reveals excellent agreement between measurement and modeling.  

\begin{figure}
\centering
\includegraphics[width=\textwidth]{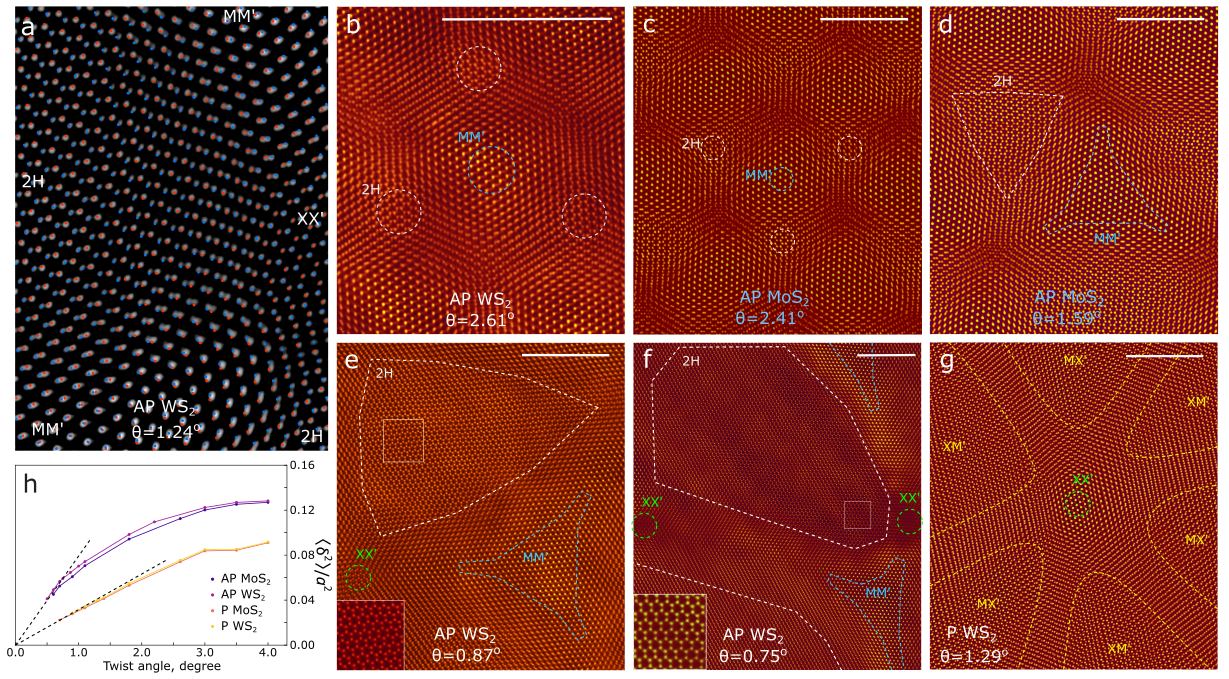}
\caption{\label{fig:III} \textbf{Evolution of commensurate domains with twist angle for AP-WS$_2$ and AP-MoS$_2$ homo-bilayers.} (\textbf{a}) Filtered experimental ADF STEM data of AP-WS$_2$ near XX$^\prime$ point (greyscale), overlaid with the calculated positions of W atoms (blue and red denote different layers). (\textbf{b-g}) Filtered ADF STEM images of samples with different periods of moir\'e cells for AP-WS$_2$ (\textbf{b, e, f, g}) and AP-MoS$_2$ (\textbf{c, d}). Twist angles cover the range from $0.54^{\circ}$ to $2.61^{\circ}$ and are measured from each image. In (\textbf{g}) the lattice reconstruction around the XX$^\prime$ stacking point in P-WS$_2$ for a twist angle of $\theta=1.29^{\circ}$ shows corners of 6 alternating XM$^\prime$ and MX$^\prime$ domains. All scale bars are 5\,nm. (\textbf{h}) Deformation parameter averaged over the supercell area showing crossover from the strong reconstruction regime (scaling linearly with the twist angle) to a ``rigid'' moir\'e pattern at large twist angles (plateau).}
\end{figure}

Dimensional analysis of Eq.(2) suggests that lattice reconstruction in the constituent monolayers is stronger for smaller twist angles. This trend is illustrated in Fig.3b-f where we show STEM images of several studied WS$_2$ and MoS$_2$ samples with different moir\'e periods corresponding to a AP-oriented homo-bilayers twisted with $\theta = 2.61^{\circ}$, 2.41$^{\circ}$, 1.59$^{\circ}$, 0.87$^{\circ}$, and 0.75$^{\circ}$. Notably, for $\theta<1.5^{\circ}$ the  MM$^{\prime}$ domain seed does not grow larger than $\sim$\SI{5}{\nm}, even as the moir\'e period increases, which leads to the kagome-like appearance of the 2H and MM$^\prime$ stacked lattice reconstruction shown in Fig.1c,d. Furthermore, for $\theta < 1^{\circ}$ two adjacent 2H/MM$^{\prime}$ boundaries merge together to form a 2H/2H boundary characterised by a screw dislocation with the Burgers vector parallel to the zigzag direction of the domain lattice (the deformation field calculated for such a single dislocation is described in SI, S10). As a result, for the smallest twist angle, the moir\'e pattern evolves into an array of dislocations running through a 2H stacked bilayer (such as in Fig.1d) and the energy gain from growing 2H areas fully overcomes the energy cost of forming the 2H/2H boundaries. This structure qualitatively differs from the simple triangular network found in P-bilayers, Fig.1b and Fig.S10, where MX$^\prime$ and XM$^\prime$ domains both represent identical lowest energy configurations (Fig.2b). The domain walls between MX$^\prime$ and XM$^\prime$ are partial screw dislocations with Burgers vector $a/\sqrt{3}$ parallel to the armchair direction of the domain lattice, that converge to points of XX$^\prime$ stacking (Fig.3g). The same trend was observed in hetero-bilayers (see Fig.1a and SI Fig.S8). 

For both orientations the transition from almost rigid bilayers to fully developed domain structures can be traced using a ``deformation parameter'', determined as the mean square of the in-plane shift, $\bm{\delta}$, between the metal and sulfur atoms in the top and bottom layers ($\bm{\delta}=0$ for perfect 2H and 3R stacking), averaged over one supercell for AP orientation and two triangular-shaped half supercells for P orientation. As $\bm{\delta}=0$ inside perfectly stacked domains, the main contribution to $\langle\bm{\delta}^2\rangle$ comes from domain boundaries (dislocations) so that it scales as the reciprocal of superlattice period, $\ell^{-1}$. The theoretically computed $\langle\bm{\delta}^2\rangle/a^2$, Fig.3h, shows a clear  $\langle\bm{\delta}^2\rangle/a^2\propto\theta$ trend at small angles, signaling formation of ideal stacking domains, and a trend towards saturation starting at $\theta_{P}^{*}\approx 2^{\circ}$ ($\theta_{AP}^{*}\approx0.9^{\circ}$) which we identify as the critical angles for formation of the dislocation network. Note that such dislocation networks may temporally evolve by dislocations reaching the sample edges, however, such recrystallization is slow,  and, here, the twist angle (which determines the network period) is pinned by clamping the crystal to the grid or a substrate.

\begin{figure}
\centering
\includegraphics[width=\textwidth]{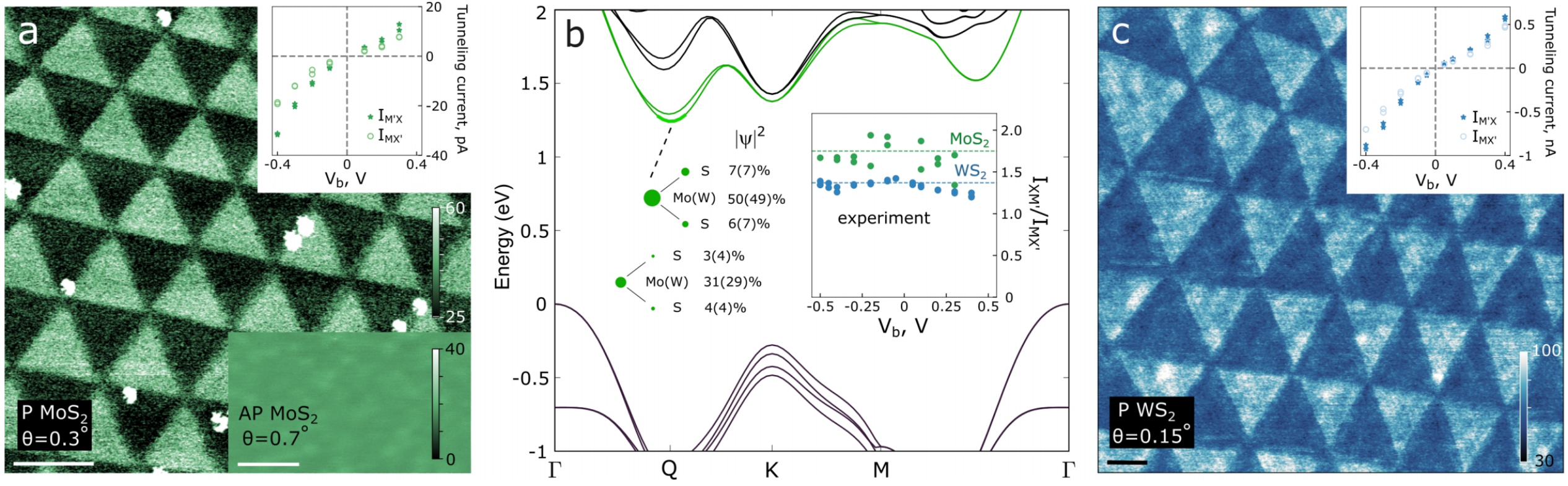}
\caption{\label{fig:IV} \textbf{Electronic properties of twisted bilayers} (\textbf{a}) CAFM map of the tunneling current acquired on a P-MoS$_2$ homo-bilayer with tip-sample bias of 600 mV at room temperature. Upper inset shows tunneling currents averaged within XM$^\prime$ and MX$^\prime$ domains respectively as a function of $V_b$. Lower inset shows CAFM mapping of AP-type MoS$_2$ homo-bilayer using the same tip at $V_b$=500 mV. (\textbf{b}) DFT bands of 3R bilayer MoS$_2$. Schematic shows contributions to orbital decomposition from molybdenum and sulphur atoms, atom sizes show qualitatively the asymmetry in the conduction band-edge wavefunctions. Values in parentheses are for WS$_2$. Inset shows experimental ratio of currents for XM$^\prime$ and MX$^\prime$ as a function of applied bias. (\textbf{c}) CAFM map of the tunneling current acquired on P-WS$_2$ homobilayer with ti p-sample bias of 50 mV at room temperature. Scale bars are 50 nm.}
\end{figure}

The presented analysis demonstrates that, below critical angles, twisted homo- and hetero-bilayers with a small lattice mismatch undergo a strong structural reconstruction into large area equilibrium domains separated by a network of dislocations. Such reconstruction necessarily changes electronic and optical properties of the twisted bilayers, as compared to the previously considered rigidly rotated crystalline lattices \cite{Seyler2019,Tran2019,Jin2019,Alexeev2019}. In particular, for $\theta<\theta_{P/AP}^{*}$ the properties of a bilayer sheet would be dominated by intrinsic properties of 2H/3R domains. For instance, this enables us to make the first measurement of the Angle-Resolved Photoemission Spectra (ARPES) of the 3R phases of MoS$_2$ and WS$_2$ (see SI S8): while those have valence band dispersions qualitatively similar to 2H stacked bilayers, we notice the difference in atomic structure through the variation of ARPES intensity across the Brillouin zone. Interestingly, the most pronounced difference between 3R and 2H stacking for bilayers, suggested by symmetry and confirmed by DFT calculations, appears to be for electrons near the conduction band edge. While the inversion symmetry of 2H stacking makes the electron wavefunction density $\lvert \psi \rvert ^2$ equal on both constituent monolayers, the lack of both inversion and mirror symmetry in 3R stacking allows  $\lvert \psi \rvert^2$ to be different on the same atoms in the top and bottom layers. 

We tested such layer polarization of bilayer bands by performing conductive atomic force microscopy (CAFM) of P-MoS$_2$ homo-bilayers, Fig.4. The results shown in these images demonstrate significantly higher (up to 2 times) tunneling current for XM$^\prime$ as compared to their MX$^\prime$ twins. The top-right inset in Fig.4a shows that such an asymmetry persists in the tip-sample bias range of $V_b = -0.5$ to $0.4V$, for which the Fermi level in the graphite substrate is close to the conduction band edge in bilayer (see SI S8). The conduction band edge appears at the Q-point of the Brillouin Zone, and DFT analysis of the orbital composition of the band edge states shows a $\sim$1.5-1.8 ratio between the weight of the electron wavefunction on the equivalent atoms in the top and bottom layer, illustrated in Fig.4b. Note that in CAFM, tunneling from the tip into XM$^\prime$ and MX$^\prime$ domains tests this distribution from the top and bottom side. For the P-orientation of twisted MoS$_2$ and WS$_2$ homo-bilayers the measured current ratio $I_{\textrm{XM}^\prime} / I_{\textrm{MX}^\prime} \approx 1.7$ (inset Fig.4b) falls very close to the theoretically estimated values, whereas for AP homo-bilayers the CAFM mapping (lower inset Fig4a) does not reveal any pronounced pattern as expected.          

Compared to twisted graphene structures, we show that twisted homo- and hetero-bilayers of TMDs offer a broader diversity of physical properties prescribed by the lack of inversion symmetry in the constituent layers. In particular, we find that lattice reconstruction for P-oriented bilayers creates condition for the formation of sizeable domains of 3R stacking (which is rarely found in bulk MoS$_2$ and WS$_2$ crystals) and features intrinsic asymmetry of electronic wavefunctions which can be used for electric tuning of the optical properties of such twisted bilayers by a displacement field. For both P and AP-stacking, the formation of dislocation networks may also result in interesting properties due to the strong deformation fields produced at domain walls and their intersections (see SI S11 for pseudo-magnetic fields calculations).

\section*{Methods}

\subsection*{STEM imaging and analysis}
Low magnification scanning transmission electron microscope (STEM) images of the twisted bilayer moir\'e patterns for different twist angles were acquired on a FEI Talos X-FEG S/TEM operated at 80 kV. High contrast moir\'e features were achieved with a convergence angle of 6 mrad and an annular dark field (ADF) collection angle range of 14-85 mrad. High resolution ADF STEM imaging was performed using a FEI Titan G2 80-200 S/TEM ChemiSTEM microscope and a JEOL ARM300CF double aberration corrected microscope with a cold FEG electron source. The Titan was operated at 200 kV with a probe current of 40 pA, a 21 mrad convergence angle, a 48-191 mrad ADF collection angle and all aberrations up to 3rd order corrected to better than a $\pi/4$ phase shift at 20 mrad. The JEOL ARM300CF was operated at an accelerating voltage of 80 kV, with a beam convergence semi-angle of 31 mrad, probe current of 10 pA and a ADF collection angle of 68-206 mrad. All aberrations were individually corrected to better than a $\pi/4$ phase shift at 30 mrad. High resolution multislice ADF image simulations were performed with QSTEM \cite{KochThesis}, matching the experimental parameters for high resolution imaging using the FEI Titan microscope, using an effective source size of \SI{1.0}{\angstrom}.
Image filtering was performed using a patch based principal component analysis (PCA) denoising algorithm \cite{Salmon2014} implemented using the open source python package Hyperspy \cite{HyperSpy}. Briefly, small patches of the image, centred on local bright peaks, were compiled to form a 3D image stack. PCA denoising was then performed on the image stack, before the denoised patches were averaged at their original locations in the image to generate a denoised image. A high-pass filter was then applied to the image to remove long range signals associated with local surface contamination. See SI for details.
\subsection*{Calculations}
For DFT calculations of the interlayer binding energy densities we used van der Waals density functional theory (vdW-DFT) with optB88 functional implemented in Quantum Espresso \cite{giannozzi2009}. In these calculations we neglected spin-orbit coupling, used a plane-wave cutoff of 816.34 eV (60 Ry), and kept the monolayer structure rigid varying interlayer distances $d$ and stacking configurations.
Details concerning the multiscale modeling can be found in the Supplementary Materials.
The density functional theory (DFT) bands were computed in the local density approximation, as implemented in the VASP code \cite{Kresse1996}, with spin-orbit coupling taken into account using projector augmented wave pseudopotentials. The cutoff energy for the plane-waves is set to 600 eV, and the in-plane Brillouin zone is sampled by a 12x12 grid. The bilayers are placed in a periodic three-dimensional box with a separation of 20 angstroms between repeated images to ensure no interaction would occur between them. The structural parameters were taken from experiments \cite{Schutte1987}.
\subsection*{CAFM imaging and analysis}
Conductive atomic force microscopy (CAFM) was performed on twisted bilayer samples using an Asylum Research Cypher-S AFM in ambient conditions mounted to steel disks using silver conductive paint. Twisted bilayers have been transferred onto multilayer graphite pre-exfoliated onto oxidized Si wafer coated with 5 nm of platinum, which served as the drain for tunneling current. CAFM images were acquired using Budget Sensors ElectriMulti75-G CAFM probes with a nominal force constant of 3 N/m, a nominal resonant frequency of 75 kHz and a conductive coating of 5 nm Cr and 25 nm Pt. Inital imaging prior to contact-mode CAFM was performed using amplitude-modulated tapping mode (AC-mode) AFM at an amplitude setpoint of 60-80 \% of the free-air amplitude when driven at 5 \% below the resonant frequency of the cantilever. Typical CAFM imaging force setpoints were between 10-40 nN and the applied bias was typically kept below $\pm$ 0.75 V during imaging. For full CAFM experimental details see supplementary information. All AFM data analysis was performed using the open-source Gwyddion software package \cite{gwyddion}.
\section*{Acknowledgments}
We acknowledge support from EPSRC grants EP/N010345/1, EP/P009050/1, EP/S019367/1, EP/S030719/1, EP/P01139X/1, EP/R513374/1 and the CDT Graphene-NOWNANO, and the EPSRC Doctoral Prize Fellowship. In addition, we acknowledge support from the European Graphene Flagship Project (696656), European Quantum Technology Flagship Project 2D-SIPC (820378), ERC Synergy Grant Hetero2D, ERC Starter grant EvoluTEM, Royal Society and Lloyd Register Foundation Nanotechnology grant. VE (reconstruction simulations) acknowledges support of Russian Science Foundation (project no. 16-12-10411).
PHB acknowledges support from the Leverhulme Trust (Research Fellowship Grant RF-2019-460).

Additional data related to this paper and the computer code used for the image filtering may be requested from the authors.

\end{document}